\documentclass[prl,aps,twocolumn,superscriptaddress,floatfix,showpacs,10pt]{revtex4-1}
\usepackage{graphicx,amsfonts,amssymb,amsmath,hyperref,dsfont,braket}
\usepackage{nicefrac}
\usepackage{multirow}

\newif\ifhyper
\hypertrue
\ifhyper
\hypersetup{
   citecolor = {green},
   colorlinks = {true}, 
   urlcolor = {blue} 
}
\fi
\newcommand{\Cov}{\mathrm{Cov}}
\newcommand{\Var}{\mathrm{Var}}

\newcommand{\beq}{\begin{equation}}
\newcommand{\eeq}{\end{equation}}
\newcommand{\beqa}{\begin{eqnarray}}
\newcommand{\eeqa}{\end{eqnarray}}

\def\ket#1{\vert#1\rangle}

\def\Longarrow{\protect\@lra}
\def\@lra{\relbar\joinrel\relbar\joinrel\relbar\joinrel%
          \relbar\joinrel\rightarrow}

\def\Im {\mbox{Im}}
\def\be{\begin{equation}}       \def\ee{\end{equation}}
\def\bea{\begin{eqnarray}}      \def\eea{\end{eqnarray}}
\def\bes{\begin{subequations}}  \def\ees{\end{subequations}}

\def\hc{\text{h.c.}}

\def\munu {\mu^{\phantom{\dagger}}_\nu}
\def\snu{\overline{s}^{\phantom{\dagger}}_\nu}
\def\snsq{\overline{s}^2_\nu}

\begin{document}

\title{Dynamic structure factor of disordered quantum spin ladders}

\author{Max H\"ormann}
\email{max.hoermann@fau.de}
\affiliation{Institute for Theoretical Physics, FAU Erlangen-N\"urnberg, Germany}

\author{Paul Wunderlich}
\email{paul.wunderlich@fau.de}
\affiliation{Institute for Theoretical Physics, FAU Erlangen-N\"urnberg, Germany}

\author{K.~P.~Schmidt}
\email{kai.phillip.schmidt@fau.de}
\affiliation{Institute for Theoretical Physics, FAU Erlangen-N\"urnberg, Germany}

\date{\rm\today}

\begin{abstract}
  We investigate the impact of quenched disorder on the dynamical correlation functions of two-leg quantum spin ladders. Perturbative continuous unitary transformations with the help of white graphs and bond-operator mean-field theory are used to calculate the one- and two-triplon contribution of the zero-temperature dynamical structure factor. Disorder results in huge effects on quasi-particles as well as composite bound states due to localization. This leads to intriguing quantum structures in dynamical correlation functions well observable in spectroscopic experiments.      
\end{abstract}


\maketitle

Disorder is an inevitable ingredient of any condensed matter. On the one hand disorder can change or even destroy the physical behaviour of the associated clean systems \cite{Vojta06,Vojta10,Griffith69} or, on the other hand, it can induce fundamentally new physics. This is especially true for correlated quantum materials where the interplay of disorder and quantum fluctuations can result in technological challenges or exotic phases of quantum matter like many-body localization \cite{Basko06,Oganesyan07,Znidaric08,Pal08}. One important aspect in the collective behaviour of correlated quantum matter is the formation of quasi-particles and their role in quantum critical behaviour. While many studies have investigated the static and thermodynamic properties of such systems in the presence of disorder \cite{Furusaki94,Westerberg95,Nagaosa96,Miyazaki97,Greven98,Melin02,Hoyos04,Trinh12,Trinh13}, the fate of quasi-particles under disorder is only rarely studied. Experimentally, however, increasingly improving resolution in spectroscopy like inelastic neutron or light scattering as well as intentional doping to control disorder in quantum materials \cite{Manaka08,Manaka09,Hong10,Stone11,huevonen12,huevonen12b,nafradi13}, demands theoretical predictions for dynamical correlation functions of correlated quantum matter in the presence of disorder.

An outstanding arena, both experimentally and theoretically, to investigate the influence of quenched disorder on quasi-particle excitations are low-dimensional quantum magnets, which host a variety of interesting quantum phases and elementary particles in the clean case. Theoretically, the calculation of dynamical correlation functions for disordered quantum magnets is challenging \cite{motrunich2001dynamics,Vojta13,shu18}. Only recently \cite{Vojta13}, the effect of disorder on single quasi-particles has been studied within bond-operator mean-field (MF) theory for the bilayer square lattice Heisenberg model, but the fate of quasi-particles under disorder beyond MF theory is largely unexplored. 

%
\begin{figure}[t]
	\centering
		\includegraphics[width=0.8\columnwidth]{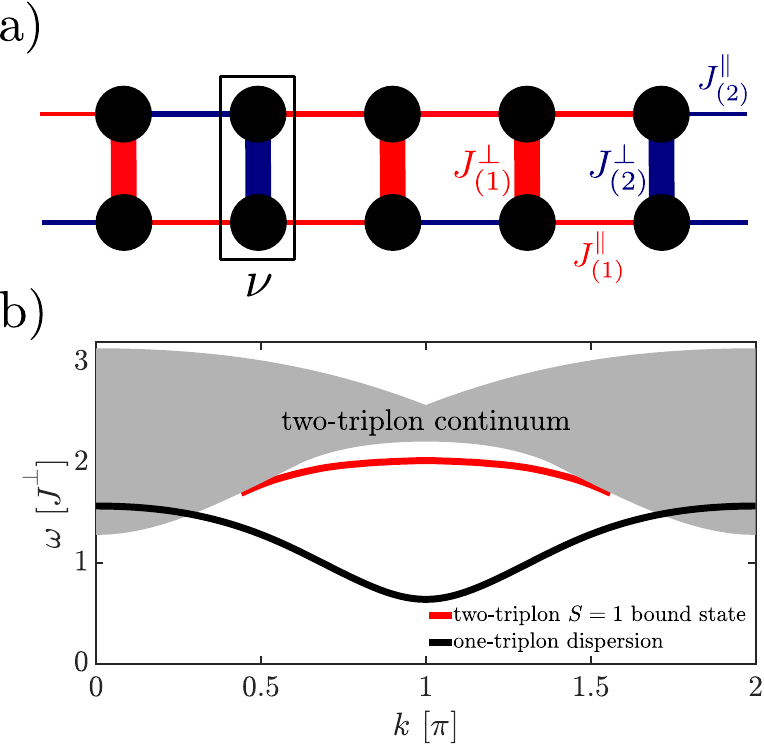}
                \caption{(a) QSL with bimodal disorder. The two different rung ($J^\perp_{(1,2)}$) and leg ($J^\parallel_{(1,2)}$) couplings are shown as red and blue bonds. The unit cell $\nu$ is indicated by the rectangular box. (b) Spectrum $\omega$ of the clean QSL for $J^\parallel/J^\perp=1/2$ as a function of momentum $k$ including one-triplon dispersion (black line), two-triplon continuum (grey) as well as spin-one two-triplon bound state (red line) calculated by pCUTs.}
	\label{fig:model}
\end{figure}

An particularly promising system to advance are antiferromagnetic two-leg quantum spin ladders (QSLs) \cite{dagotto96}, which have been investigated successfully over the last decades in a number of experimental realizations \cite{Windt01,Notbohm07,Ruegg08b,Hong10b,Ward13,Schmidinger13,Ward17} and by various theoretical tools \cite{Barnes93,Shelton96,Trebst00,Knetter01,Schmidt01,Lauchli03}. Clean QSLs have non-ordered ground states and gapped triplon excitations \cite{Shelton96,Schmidt03c}. Inelastic neutron and light scattering allow to access one-triplon dispersions but also the formation of two-triplon bound states and continua reflecting the presence of large quantum fluctuations. Furthermore, it is possible experimentally to intentially dope QSL compounds so that quenched disorder is induced into the system \cite{WardPhD}. It is therefore important to understand the influence of disorder on the spectral signatures of triplon quasi-particles in QSLs. This is exactly the punchline of this letter. 
We calculate the one- and two-triplon contribution of the dynamic structure factor (DSF) at $T=0$ of two-leg QSLs in the presence of quenched disorder.  It is demonstrated that disorder has huge effects on collective excitations yielding intriguing quantum structures directly relevant for spectroscopic experiments.    

%
%
\emph{Disordered QSL ---} 
%
%
The Hamiltonian of the disordered QSL for a fixed disorder configuration $\{J\}$ is given by
%
\begin{equation}
 \mathcal{H}\left(\{J\}\right) = \sum_\nu\left( J_{\nu}^{\perp}\;\vec{S}_{\nu,1}\cdot \vec{S}_{\nu,2}+\sum_{n=1}^2 J_{\nu,n}^{\parallel}\;\vec{S}_{\nu,n}\cdot \vec{S}_{\nu+1,n}\right)\, , 
\label{Eq::Ham_QSL}
\end{equation}
%
where the sum runs over all rungs and $n=1,2$ denotes the legs of the QSL (see Fig.~\ref{fig:model}a). The disorder configuration $\{J\}$ given by the antiferromagnetic $J_{\nu}^{\perp}$ and $J_{\nu,n}^{\parallel}$ depends on the type of quenched disorder. Here we focus on bimodal disorder, i.e.~the rung and leg exchanges can take either the value $J_{1}^{\kappa}$ with probability $p$ or $J_{2}^{\kappa}$ with probability $1-p$ for $\kappa\in\{\perp,\parallel\}$. However, our technical treatment is more general and allows to consider any form of quenched bond disorder.

Clean QSLs with $J_{\nu}^{\perp}\equiv J^{\perp}$ and $J_{\nu,n}^{\parallel}\equiv J^{\parallel}$ have an unordered ground state and gapped triplon excitations for all $J^{\parallel}/J^{\perp}$, which are adiabatically connected to the isolated rung limit $J^{\parallel}=0$. In this limit the ground state is a product state of singlets \mbox{$|s\rangle$=$(|\uparrow\downarrow\rangle - |\downarrow\uparrow\rangle )/\sqrt{2}$} and excitations are local triplets $|t^{+1}\rangle$=$|\uparrow\uparrow\rangle$, \mbox{$|t^{+0}\rangle$=$(|\uparrow\downarrow\rangle + |\downarrow\uparrow\rangle )/\sqrt{2}$}, and \mbox{$|t^{-1}\rangle$=$|\downarrow\downarrow\rangle$}. The low-energy spectrum of the clean QSL obtained by pCUTs is shown for $J^{\parallel}/J^{\perp}=1/2$ in Fig.~\ref{fig:model}b. The spin-one triplon corresponds to a dressed triplet which has a finite cosine-like dispersion in this parameter regime with a finite gap at $k=\pi$. Two-triplon energies are either part of the two-triplon continuum or correspond to two-triplon (anti)-bound states, whose energies depend on the total spin. We focus on the spin-one sector, which is relevant for the DSF, where a two-triplon bound state exists for a large range of momenta in the clean QSL due to the attractive two-triplon interaction \cite{Trebst00,Knetter01}.      

We then reformulate \eqref{Eq::Ham_QSL} in terms of triplet creation and annihilation operators $t^{(\dagger )}_{\nu,\alpha}$ with $t^{\dagger}_{\nu,\alpha }|s\rangle \equiv |t^{\alpha}\rangle $ and \mbox{$\alpha\in\{\pm 1,0\}$} on rung $\nu$. Setting the average rung exchange \mbox{$\bar{J}^\perp\equiv (J_{1}^{\perp} +J_{2}^{\perp})/2\equiv 1$} and introducing the deviations $\Delta\bar{J}^\perp_{\pm}$ from it, allows to express \eqref{Eq::Ham_QSL} as
%
\begin{equation}
 \mathcal{H}\left(\{J\}\right) = E_0+\mathcal{Q}+\sum_{n=-2}^2 \hat{T}_n\left(\{J\}\right)\quad , 
\label{Eq::Ham_QSL_Trp}
\end{equation}
%
where $E_0=-3 \mathcal{N}_{\rm r}/4$ with $\mathcal{N}_{\rm r}$ the number of rungs, the counting operator \mbox{$\mathcal{Q}=\sum_{\nu,\alpha}\hat{n}_{\nu,\alpha}$} with $\hat{n}_{\nu,\alpha}=t^\dagger_{\nu,\alpha}t^{\phantom{\dagger}}_{\nu,\alpha}$, and the \mbox{$\hat{T}_n$} with $[\hat{T}_n,\mathcal{Q}]=n\hat{T}_n$ change the triplet number by $n$. The $\hat{T}_n$ depend explicitly on $\Delta\bar{J}^\perp_{\pm}$ as well as $J^\parallel_{1,2}$ and therefore on the disorder configuration $\{J\}$. Here $\hat{T}_{\pm 2}$ correspond to pair creation and annihilation processes, $\hat{T}_{0}$ contains triplet hopping as well as quartic triplet-triplet interactions, and $\hat{T}_{\pm 1}$ represent decay processes of one triplet into two or vice versa. Note that $\hat{T}_{\pm 1}=0$ holds for the clean case where the QSL possesses an exact reflection symmetry about the centerline giving rise to a conserved parity quantum number $\pm 1$.

The central quantity for inelastic neutron scattering on disordered QSLs is the disorder averaged DSF
%
\begin{equation}
 S_{\pm}(k,\omega) \equiv \lim_{\mathcal{N}_{\rm dc}\rightarrow\infty}\frac{1}{\mathcal{N}_{\rm dc}} S_\pm (k,\omega,\{J\}) 
\label{Eq::DSF_Averaged}
\end{equation}
%
with momentum $k$, frequency $\omega$, number of disorder configurations $\mathcal{N}_{\rm dc}$, and 
%
\begin{equation}
 S_{\pm}(k,\omega,\{J\}) \equiv -\frac{1}{\pi}{\rm Im}\langle 0| \mathcal{O}^{\dagger}_\pm \frac{1}{\omega-\mathcal{H}+{\rm i}0^+} \mathcal{O}^{\phantom{\dagger}}_\pm|0\rangle\label{Eq::DSF}
\end{equation}
%
where $\mathcal{O}_\pm(k)\equiv\sum_\nu e^{{\rm i}k\nu} (S_{\nu,1}^z\pm S_{\nu,2}^z)/(2\sqrt{\mathcal{N}_r})$. The index $\pm $ reduces to the parity quantum number for the clean QSL.  

%
%
%
%

%
\begin{figure*}[t]
	\centering
		\includegraphics[width=0.8\textwidth]{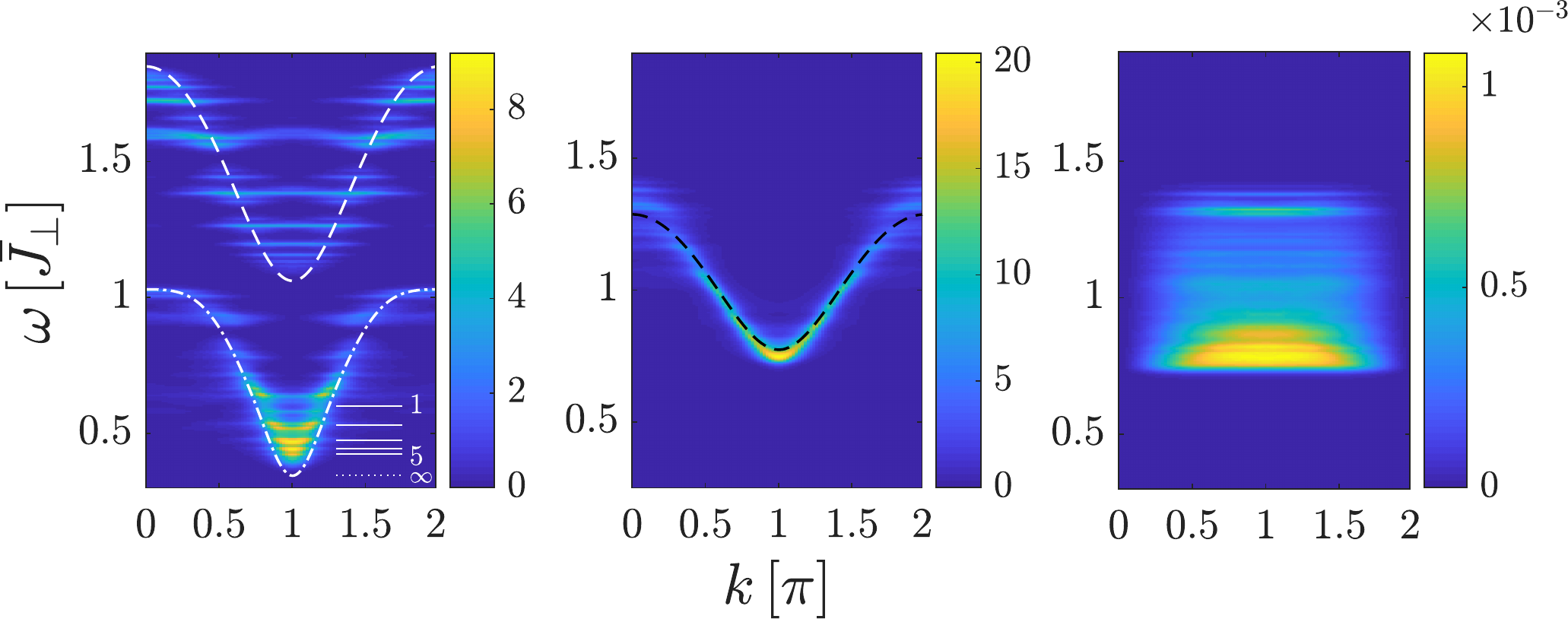}
                \caption{One-triplon contribution to the anti-symmetric DSF $S_{-}(k,\omega)$ for pure rung (left) or leg (middle) disorder and to the symmetric DSF $S_{+}(k,\omega)$ for pure leg disorder (right).
	        {\it Left}: bimodal rung disorder with $p=0.5$ and rung exchanges $J_{1}^{\perp}=1.4$ and $J_{2}^{\perp}=0.6$. The constant leg exchange is $J^{\parallel}=0.4$. White lines represent one-triplon dispersions for a clean QSL with $J^{\parallel}=0.4$ and $J^{\perp}=1.4$ (dashed) or $J^{\perp}=0.6$ (dashed-dotted). Horizontal white lines indicate the lowest one-triplon energy $\epsilon_1^{(L)}$ of open ladder segments of length $L$ with $J^{\perp}=0.6$.           
		{\it Middle/Right}: bimodal leg disorder with $p=0.5$ and leg exchanges $J_{1}^{\parallel}=0.1$ and $J_{2}^{\parallel}=0.4$. The black dashed line represents the one-triplon dispersion with mean hopping amplitudes.}
	\label{fig:1QP}
\end{figure*}

%
%
\emph{pCUT ---}
%
%
We perform pCUTs as our main approach as well as bond-operator MF theory (for the latter see \cite{Supp}). The major target of a pCUT is to unitarily transform \eqref{Eq::Ham_QSL_Trp}, order by order in $J_{\nu,n}^{\parallel}$ and $\Delta\bar{J}^\perp_{\pm}$, to an effective Hamiltonian $\mathcal{H}_{\rm eff}$ which conserves the number of triplons so that $[\mathcal{H}_{\rm eff},\mathcal{Q}]=0$ holds. As a consequence, the complicated quantum many-body system is mapped to an effective few-body problem which is easier to solve. A pCUT application has a model-independent step, which expresses $\mathcal{H}_{\rm eff}$ in a sum of operator product sequences of the $\hat{T}_n$ operators with exactly known coefficients. The most efficient way of performing the second model-dependent step, which amounts to normal order $\mathcal{H}_{\rm eff}$, is a full-graph decomposition using the linked-cluster theorem. Here the only graphs are ladders segments so that the calculation for the clean QSL is very simple. In contrast, in the presence of bimodal disorder, there are $2^{3\mathcal{N}_r+1}$ different graphs for a ladder segment of $\mathcal{N}_r$ rungs and a linked-cluster expansion becomes unefficient since $\mathcal{N}_{\rm dc}$ is large. At this point white graphs \cite{Coester15} are essential, since it allows to specify $\{J\}$ only {\it after} the calculations on the graphs. In practice, one determines the most general linked contribution of a graph by allowing for a different exchange coupling on every nearest-neighbor link of the graph. The resulting multi-variable series can then be embedded on any specific $\{J\}$. 

We calculated $\mathcal{H}_{\rm eff}$ in the one- and two-triplon sector up to order $8$ and we determined the corresponding effective observables $\mathcal{O}^{\rm eff}_\pm$ up to order $7$. The convergence of the bare series is similar to the clean QSL where it gives quantitative results up to $J^\parallel/J^\perp\lesssim 0.5$ \cite{Knetter01,Schmidt01} and we therefore restrict to this parameter regime below. The effective one- and two-triplon problem is then diagonalized for finite QSLs with $\mathcal{N}_{\rm r}=100$ and the DSF for a fixed disorder configuration is obtained using a finite broadening $\Gamma=0.01$. Averaging over $\mathcal{N}_{\rm dc}=1000$ disorder configurations then gives the averaged DSF \eqref{Eq::DSF_Averaged}.  All relevant aspects discussed below are well converged despite the use of finite $\mathcal{N}_{\rm r}$ and $\mathcal{N}_{\rm dc}$ \cite{Supp}, which is reasonable due to the finite localization and correlation length of the QSL.

\emph{One-triplon contribution ---}
%
%
For the clean QSL, this sector can be expressed as \mbox{$S_{-}(k,\omega)=a^2(k)\;\delta(\omega-\omega_{k})$} with the one-triplon dispersion $\omega_{k}$ and the one-triplon spectral weights $a^2(k)$ while $S_{+}(k,\omega )=0$ due to the parity symmetry. The dispersion $\omega_{k}$ is cosine shaped with gap at $k=\pi$ and the spectral weight $a^2(k)$ is monotonic in $k$ with maximum at $k=\pi$ and minimum at $k=0$.

Representative pCUT results for the DSF in the presence of maximal pure rung or leg disorder are shown in Fig.~\ref{fig:1QP} (see \cite{Supp} for qualitatively the same MF results). These two cases behave differently with respect to reflection about the centerline. For rung disorder the associated parity is still a conserved quantity and therefore $S_{+}(k,\omega )=0$ holds for the one-triplon sector. In contrast, since the leg disorder breaks the reflection symmetry, a finite one-triplon contribution to $S_{+}(k,\omega )$ exists. However, this contribution is orders of magnitude smaller and the dominant contribution to the DSF is still $S_{-}(k,\omega )$.
We find that the DSF of the disordered QSLs is fundamentally different to the one for the clean counterpart. These differences do not originate from large changes of the spectral weights. Indeed, deviations to a clean QSL with constant mean exchange couplings are small \cite{Supp}. Certainly, the most important effect of disorder on the DSF is the disorder-induced localization of eigenfunctions and their corresponding shape changes in momentum space. Furthermore, the density of states is changed by the bimodal disorder \cite{Supp}.

There are fundamental differences between leg and rung disorder, although eigenfunctions are localized in both cases. This is understood in leading-order perturbation theory: rung disorder leads to triplons hopping in a disordered chemical potential while leg disorder yields a disordered nearest-neighbor hopping and therefore to a momentum-dependent disorder potential. Thus localization is stronger for rung compared to leg disorder.

We find two separated energy regions for rung disorder corresponding to $J_{1}^{\perp}$ and $J_{2}^{\perp}$. This can be seen by comparing to the one-triplon dispersions of a clean QSL with these $J_{\nu}^{\perp}$ (white lines in Fig.~\ref{fig:1QP}). In first order this follows from Gerschgorin's circle theorem which states that half of the eigenvalues are bigger and half of them smaller than one in the thermodynamic limit for that disorder.

%
\begin{figure*}[t]
	\centering
	\includegraphics[width=0.8\textwidth]{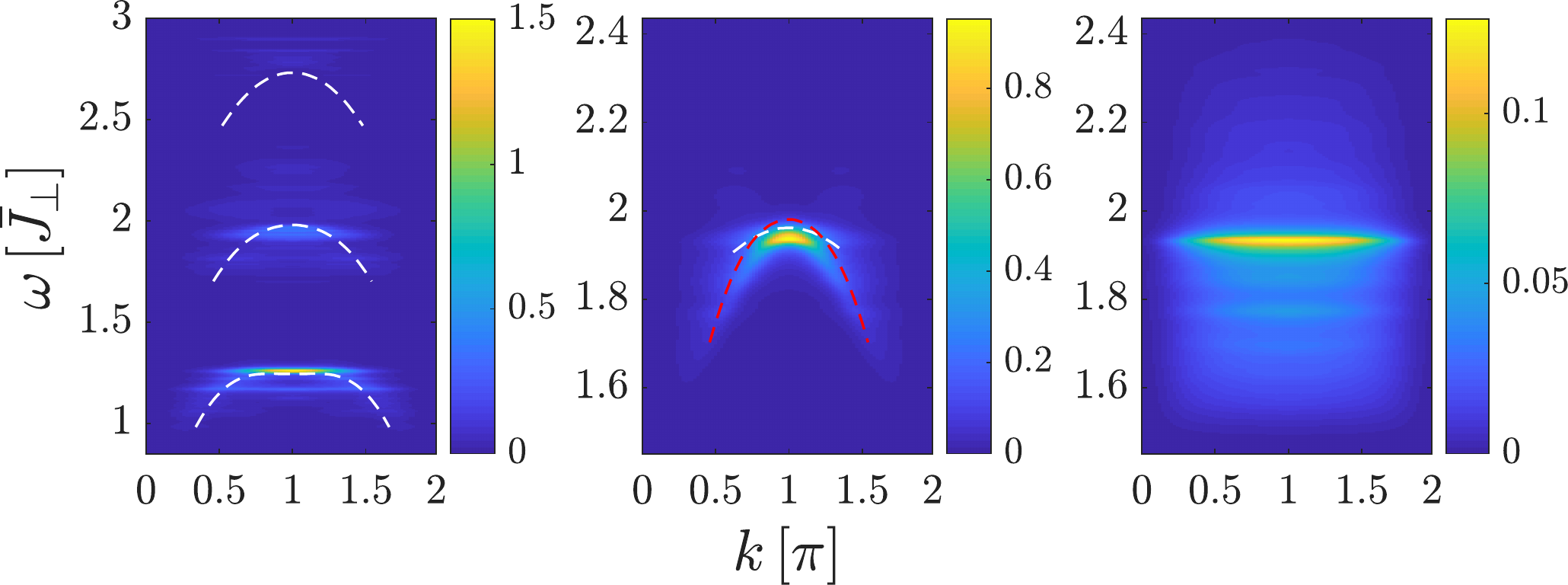}
	\caption{Two-triplon contribution to $S_{+}(k,\omega)$ for pure rung (left) or leg (middle) disorder and to $S_{-}(k,\omega)$ for pure leg disorder (right).
 {\it Left}: bimodal rung disorder with $p=0.5$ and rung exchanges $J_{1}^{\perp}=1.4$ and $J_{2}^{\perp}=0.6$. The constant leg exchange is $J^{\parallel}=0.4$. The three dashed lines represent the dispersion of the two-triplon bound state for the clean QSL using $J^{\perp}=1.4$,  $J^{\perp}=1.0$, and $J^{\perp}=0.6$.         {\it Middle/Right}: bimodal leg disorder with $p=0.5$ and leg exchanges $J_{1}^{\parallel}=0.1$ and $J_{2}^{\parallel}=0.4$. The dashed lines represent the dispersion of the two-triplon bound state for the clean QSL with $J_{1}^{\parallel}$ (white) and $J_{2}^{\parallel}$ (red).}
	\label{fig:2QP}
\end{figure*}
%
%

The spectral weights are higher in the low-energy region due to localization. Considering an eigenstate in the lower (higher) region, it will be localized around states with dominantly low (high) rung values. For these eigenstates the average ratio of leg and rung coupling is then larger at lower energies. Hence the spectral weight is larger in the low-energy region. Most importantly, we observe a fragmentation of the one-triplon DSF into energies carrying maximal spectral weights close to $k=\pi$. These energies with largest intensity are located well above the one-triplon gap of the uniform QSL with the lower rung value. These energies can be understood by considering leading-order degenerate perturbation theory for small $J^{\parallel}/\Delta\bar{J}^\perp$ \cite{Supp}. In this limit one has a fragmentation of the disordered QSL into decoupled ladder segments with constant $J^{\perp}=0.6$ of length $L$. In each open segment one can easily solve the full one-triplon problem and determine the lowest one-triplon energy $\epsilon_1^{(L)}$. For $k=\pi$, we find that the one-triplon DSF has indeed local intensity maxima at these energies (see solid white lines in Fig.~\ref{fig:1QP}). The approximate quantization of the one-triplon DSF is therefore directly linked to the discrete energies $\epsilon_1^{(L)}$ arising from the strong rung disorder. Note that similar quantum structures occur also for the high-energy region of the one-triplon DSF which can be again traced back to a fragmentation of the disordered QSL into almost decoupled ladder segments. For leg disorder one has only a single energy region which is well described with a one-triplon dispersion using mean hopping amplitudes (black line in Fig.~\ref{fig:1QP}). The spectral weight is largest at $k=\pi$ and diminishes towards $k=0$. However, the intensities are even smaller at small $k$ as one expects from the spectral weight, since it is distributed over a particularly broad range of energies at $k=0$. This comes from disorder in the local hopping amplitude (like for rung disorder) which appears in subleading orders leading to an anisotropy between $k=0$ and $k=\pi$ \cite{Supp}. Further, the fragmentation at $k=0$ is of similar origin as for rung disorder found for all momenta. Finally, the shape of the one-triplon contribution to $S_{+}(k,\omega )$ can be fully traced back to the different type of observable. Although the spectral weight vanishes exactly at $k=0$, it is distributed almost uniformly between $k=\pi/2$ and $k=\pi$ \cite{Supp}. The same is true for the intensities at fixed energy. For the latter one has to recapitulate that the local observable in real space only injects a finite weight for single triplons if an asymmetry of leg couplings is present in the local surrounding of the observable. The probability of such an asymmetric and connected region of length $L$, where $\mathcal{O}_+(k)$ is active, is exponentially descreasing with $L$. It follows that the projection of eigenstates on these regions is almost flat in momentum space.   

%
%
\emph{Two-triplon contribution ---}
%
%
For the clean QSL, it is contained solely in $S_{+}(k,\omega )$ while $S_{-}(k,\omega )=0$. It comprises a continuum and a two-triplon bound state (see Fig.~\ref{fig:model}b). By far most of the spectral weight is carried by this bound state, which therefore dominates the DSF. 

The two-triplon contributions obtained by pCUTs with maximal pure rung or leg disorder are shown in Fig.~\ref{fig:2QP}. Note that bond-operator MF does not yield satisfactory results in this sector, since the attractive triplon interaction is neglected completely \cite{Supp}. Localization also occurs for two-triplon states so that eigenstates have almost all their weight on a finite part of the two-triplon space in the position basis \cite{Supp}. Further, the states carrying most of the weight will be two-triplon states localized on a finite region of the lattice. For rung disorder, there are 3 possible values for the sum of two local hopping amplitudes resulting in 3 distinct structures in the DSF. The spectral weight of these structures decreases from lower to higher energy, which can be understood from the decreasing average ratio of leg and rung couplings as described for the one-triplon case above. Each region contains a bound state and a continuum, although only the bound state carries significant spectral weight (see Fig.~\ref{fig:2QP} left). They gain a finite lifetime due to the disorder. The same is true for leg disorder. One observes two bound-state structures, which fuse at $k=\pi$. The two structures, one more dispersive than the other, can be linked to the bound states of clean QSLs taking the larger or lower leg coupling (see Fig.~\ref{fig:2QP} middle). Interestingly, the maximal weight at $k=\pi$ is located at a lower energy compared to the bound states of the clean QSLs. This is likely caused by scattering events of lower-energy bound states with $k<\pi$ yielding a final momentum $k=\pi$. Finally, the two-triplon contribution to $S_{-}(k,\omega )$ (see Fig.~\ref{fig:2QP} right), which is induced by the leg disorder, has a much larger weight compared to the one-triplon contribution to $S_{+}(k,\omega )$ although the shape is similar. Notably, the maximum intensity is at the same energy as for $S_{+}(\pi,\omega )$ and is again associated with bound states.

%
\emph{Conclusions ---}
%
%
The pCUT method is an efficient tool to investigate the fate of quasi-particles under quenched disorder, which we exemplified by calculating the DSF of disordered QSLs. Disorder-induced quantum structures like the quantization of collective triplon excitations in the DSF emerge which are of direct relevance for spectroscopic experiments. We therefore suspect that our findings trigger experimental as well as further theoretical investigations on the fate of quasi-particles in disordered correlated quantum matter.    

\acknowledgments We thank Bruce Normand and Christian R\"uegg for fruitful discussions. MH and KPS acknowledge financial support from DFG project SCHM 2511/10-1. 

\clearpage

\widetext
%

%
\begin{center}
\textbf{\large Supplementary Materials to "Dynamic structure factor of disordered quantum spin ladders".}

Max H\"ormann, Paul Wunderlich and K.~P.~Schmidt
\end{center}

\setcounter{equation}{0}
\setcounter{figure}{0}
\setcounter{table}{0}
\setcounter{page}{1}
\makeatletter
\renewcommand{\theequation}{S\arabic{equation}}
\renewcommand{\thefigure}{S\arabic{figure}}

\thispagestyle{empty}

At first a complementary mean-field approach for the problem is discussed. Then the construction of Hamiltonian and observables with pCUTs is explained in more detail. The way the lifetime of a momentum state changes with disorder and how this depends on the momentum itself is examined by considering self-energy calculations. For rung disorder we found an approximate decoupling of the ladder into smaller ladder segments containing only one of the two rung values in the main body of the paper. This behaviour is explained with first-order degenerate perturbation theory here. For all disorder setups of the main body of the paper the spectral weights, the density of states and the inverse participation ratio is shown. Finally we check the convergence of the pCUT calculations with respect to number of rungs $\mathcal{N}_r$, number of disorder configurations $\mathcal{N}_{\rm{dc}}$, and the order of the perturbative expansion. 
\section{Mean-field approach}
\begin{figure}[htbp]
	\centering
	\includegraphics[width=0.8\columnwidth]{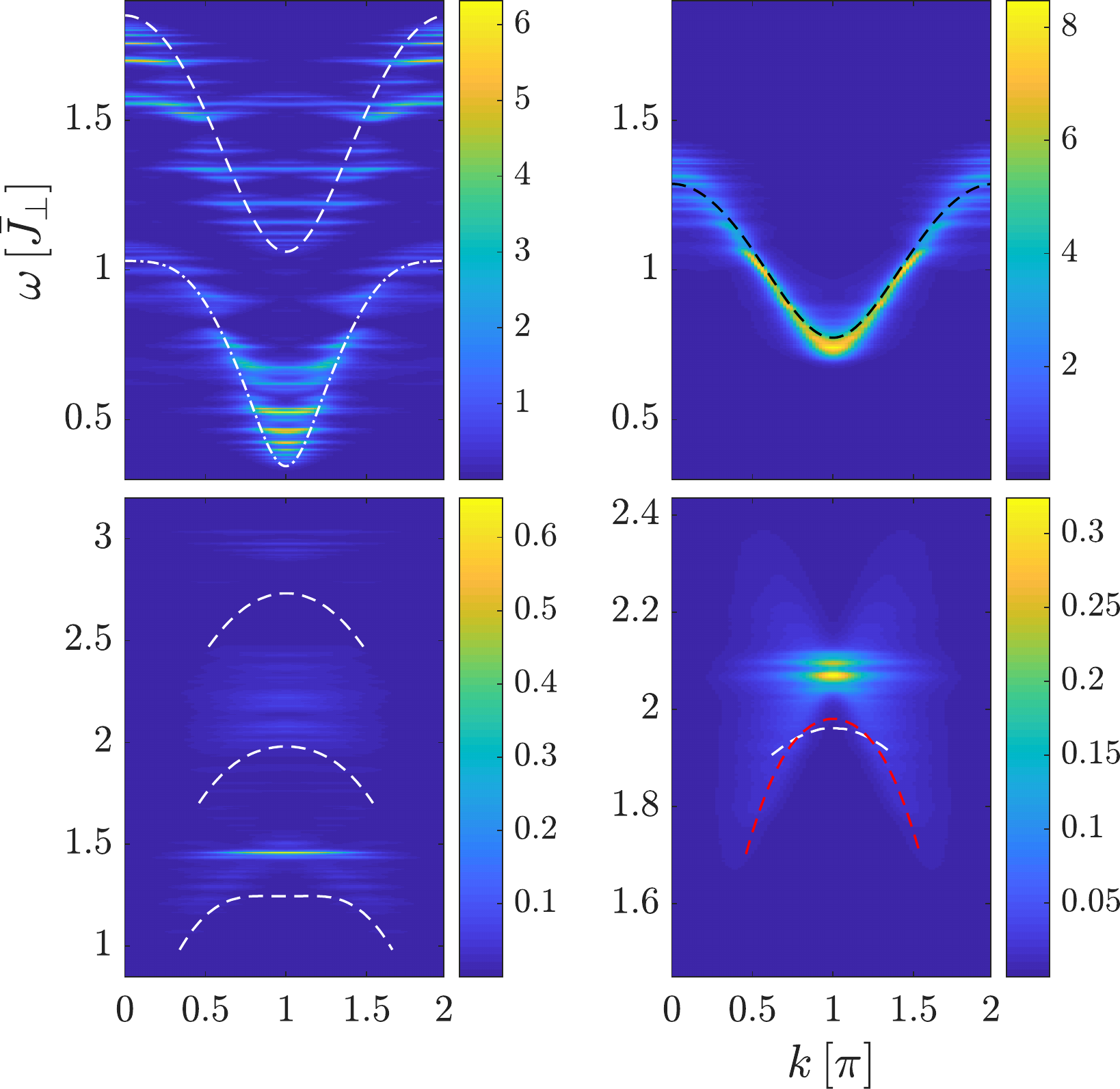}
	\caption{Uper panel: one-triplon contribution to the dynamic structure factor $S_-(k, \omega)$ using the mean-field approach for bimodal rung disorder with $p=0.5$, $J^\perp_1=1.4$, $J^\perp_2 = 0.6$ and constant leg exchange $J^\parallel = 0.4$ (\textit{left}) and bimodal leg disorder with $p=0.5$, $J_1^\parallel = 0.1$, $J_2^\parallel = 0.4$ and constant rung exchange $J^\perp=1$ (\textit{right}). White lines (\textit{left}) represent one-triplon dispersions for a clean QSL with $J^{\parallel}=0.4$ and $J^{\perp}=1.4$ (dashed) or $J^{\perp}=0.6$ (dashed-dotted). The black dashed line (\textit{right}) represents the one-triplon dispersion with mean hopping amplitudes. Lower panel: two-triplon contribution to the dynamic structure factor $S_+(k, \omega)$ using the mean-field approach for the same disorder setups as in the upper panel. The three dashed lines (\textit{left}) represent the dispersion of the two-triplon bound state for the clean QSL using $J^{\perp}=1.4$,  $J^{\perp}=1.0$, and $J^{\perp}=0.6$. In the leg disorder plot the dashed lines represent the dispersion of the two-triplon bound state for the clean QSL with $J_{1}^{\parallel}$ (white) and $J_{2}^{\parallel}$ (red). All lines were obtained by pCUTs and are the same as the ones shown in the main body of the paper. The calculations were broadened with a Lorentzian of $\Gamma=0.01$.}
	\label{DynamicSF_MF}
\end{figure}

In the framework of the mean-field approach \cite{Vojta13,Gopalan,NormandMeanField} all cubic and quartic terms of the full Hamiltonian, that was introduced in the main body of the paper in terms of triplet operators, are discarded at first. Afterwards the hardcore constraint \mbox{$s^{\dagger}_\nu s^{\phantom{\dagger}}_\nu  + \sum_\alpha t^{\dagger}_{\nu,\alpha} t^{\phantom{\dagger}}_{\nu,\alpha} =1$} is taken into account by Lagrange multipliers $\munu$ and furthermore the singlet expectation value $\braket{s^{\phantom{\dagger}}_\nu} = \snu$ is introduced. The mean-field Hamiltonian reads

\begin{equation}
\begin{aligned}
\widetilde{\mathcal{H}}^{\text{mf}}(\{J\}) = \widetilde{E}_0^{\text{mf}} + \sum_{\nu, \alpha} \left\{\left(\frac{1}{4} J_\nu^\perp -\mu_\nu\right) t^\dagger_{\nu, \alpha} t^{\phantom{\dagger}}_{\nu,\alpha} + \frac{1}{4} \sum_{n=1}^2  J_{\nu, n}^\parallel \overline{s}^{\phantom{\dagger}}_\nu \overline{s}^{\phantom{\dagger}}_{\nu+1} \left(t^{\phantom{\dagger}}_{\nu,\alpha} t^{\phantom{\dagger}}_{\nu+1,\alpha} + t^{\phantom{\dagger}}_{\nu,\alpha} t^{\dagger}_{\nu+1,\alpha} + \hc\right) \right\}
\end{aligned}
\label{MF_Hamiltonian}
\end{equation}
with $\widetilde{E}_0^{\text{mf}} = \sum_{\nu} (-\frac{3}{4} J^\perp_\nu \snsq - \munu \snsq + \munu)$. To obtain the parameters $\munu$ and $\snu$ it is necessary to solve self-consistent equations
\begin{equation}
\begin{aligned}
\Braket{\frac{\partial \mathcal{H}^{\text{mf}}(\{J\})}{\partial \munu}} = 0 \qquad \Braket{\frac{\partial \mathcal{H^\text{mf}}(\{J\})}{\partial \snu}} = 0,
\end{aligned}
\label{self_cons_eq}
\end{equation}

where $\Braket{\cdot}$ denotes the vacuum expectation value. If one tries to do so for a disordered ladder, a $2\mathcal{N}_r$ system of equations would remain which would be analytically and numerically demanding. Alternatively, self-consistent equations for ladders without disorder, that were also considered within the context of mean-field theory, are used to calculate the energy properties and then the dynamic structure factor. \\For such clean ladders the parameters $\munu$ and $\snu$ are uniform and the hardcore-constraint restricts to a global one. Starting from \eqref{MF_Hamiltonian} with $\munu = \mu$ and $\snu = \overline{s}$ a Fourier transformation was done to transform the Hamiltonian into momentum space and a Bogoliubov transformation for diagonalizing the Hamiltonian subsequently. Eventually self-consistent equations can be obtained analogue to \eqref{self_cons_eq} which can be solved numerically, so all energy properties and the dynamic structure factor of ladders without disorder can be obtained. \\These self-consistent equations for ladders without disorder are also used for ladders with disorder. To calculate $\munu$ and $\snu$, respectively, $J_\nu^\perp$ and an averaged value for $J^\parallel_\nu = \tfrac{1}{4} \sum_{n=1}^{2} (J^{\parallel}_{\nu-1, n} +J^{\parallel}_{\nu, n} )$ was used. Consider, however, if one would use global parameters $\munu = \mu$ and $\snu = \overline{s}$ for a disordered ladder, the outcome of this approach would be much worse than using local parameters because the local fluctuations of the disorder are not sufficiently taken into account. On that basis a Bogoliubov transformation of \eqref{MF_Hamiltonian} can be conducted in a real-space setting

\begin{equation}
\begin{aligned}
\mathcal{H}^{\text{mf}}(\{J\}) = E_0^{\text{mf}} + \sum_{\nu, \alpha} \omega_\nu^{\text{mf}} \gamma_{\nu,\alpha}^\dagger\gamma^{\phantom{\dagger}}_{\nu, \alpha}
\end{aligned}
\end{equation}
with the one-particle energies of the Bogoliubov quasi-particles $\omega^{\text{mf}}$ and $E_0^{\text{mf}} = \sum_{\nu}(-\frac{3}{4} J_\nu^\perp \snsq - \munu \snsq + \frac{5}{2} \munu- \frac{3}{8} J^\perp_\nu + \frac{3}{2} \omega^{\text{mf}}_\nu)$.
Starting from this result, $\omega^{\text{mf}}$ and the corresponding eigenstates can be used to calculate the dynamic structure factors $S_\pm(k, \omega)$ introduced in the paper. \\ The mean-field results for the same bimodal leg and rung disorder setups as in the main body of the paper are shown in Fig.~\ref{DynamicSF_MF}. In the one-triplon sector ($S_-(k, \omega)$), they exhibit a good qualitative accordance to the results obtained by pCUT calculations. Nevertheless some deviations in shape and in spectral weight appear by construction, but we see for coupling constants up to $J^{\parallel}_{\nu, n} / J^\perp_\nu = 0.1$ that the outcome of the mean-field approach shows nearly the same quantitative behavior as the pCUT approach. The dynamic structure factor $S_+(k, \omega)$ that contains two-triplon contributions was also calculated using mean-field theory. As expected, only the two-triplon continuum can be obtained and the contribution of bound states is missing because of the neglection of triplon interactions. Note that the mean-field approach can be applied to arbitrary disorder configurations. \\

\section{Construction of Hamiltonian and observable with pCUT}
Rung and leg couplings are randomly chosen for a finite lattice of size $\mathcal{N}_r$ according to a fixed disorder distribution. To construct the Hamiltonian one needs all matrix elements between one- and two-triplon states for that realization of the disorder. For a given order of the perturbation theory only linked subclusters with a length of at most that order contribute to these matrix elements. For all such subclusters of that particular realization the hopping amplitudes of triplons or the interactions between two triplons are calculated. Afterwards these amplitudes are properly assigned to the corresponding matrix elements between one- or two-triplon states on the finite lattice. The remaining task is to diagonalize the obtained matrix.\\
For the observable the treatment is similar. At first the local observables $\mathcal{O}_\pm(\nu)$ and their action on the ground state is calculated for every rung $\nu$ of the disorder realization. This action is not local anymore due to the pCUT transformation and the observable can create one- or two-triplon states at most distanced by the perturbation order from $\nu$. The amplitudes for the triplon states that are created by the local observables are again created for all subclusters of the lattice. The local actions have to be Fourier transformed to obtain the global action $\mathcal{O}_\pm(k)$. It is convenient to calculate the action of $\mathcal{O}_\pm(k)$ in position space. For that one has to take care to properly account for the phases caused by the non-local triplon actions of the local triplet observables.

\section{Momentum dependence of momentum state lifetimes}
\begin{figure}[htbp]
	\centering
	\includegraphics[width=0.9\columnwidth]{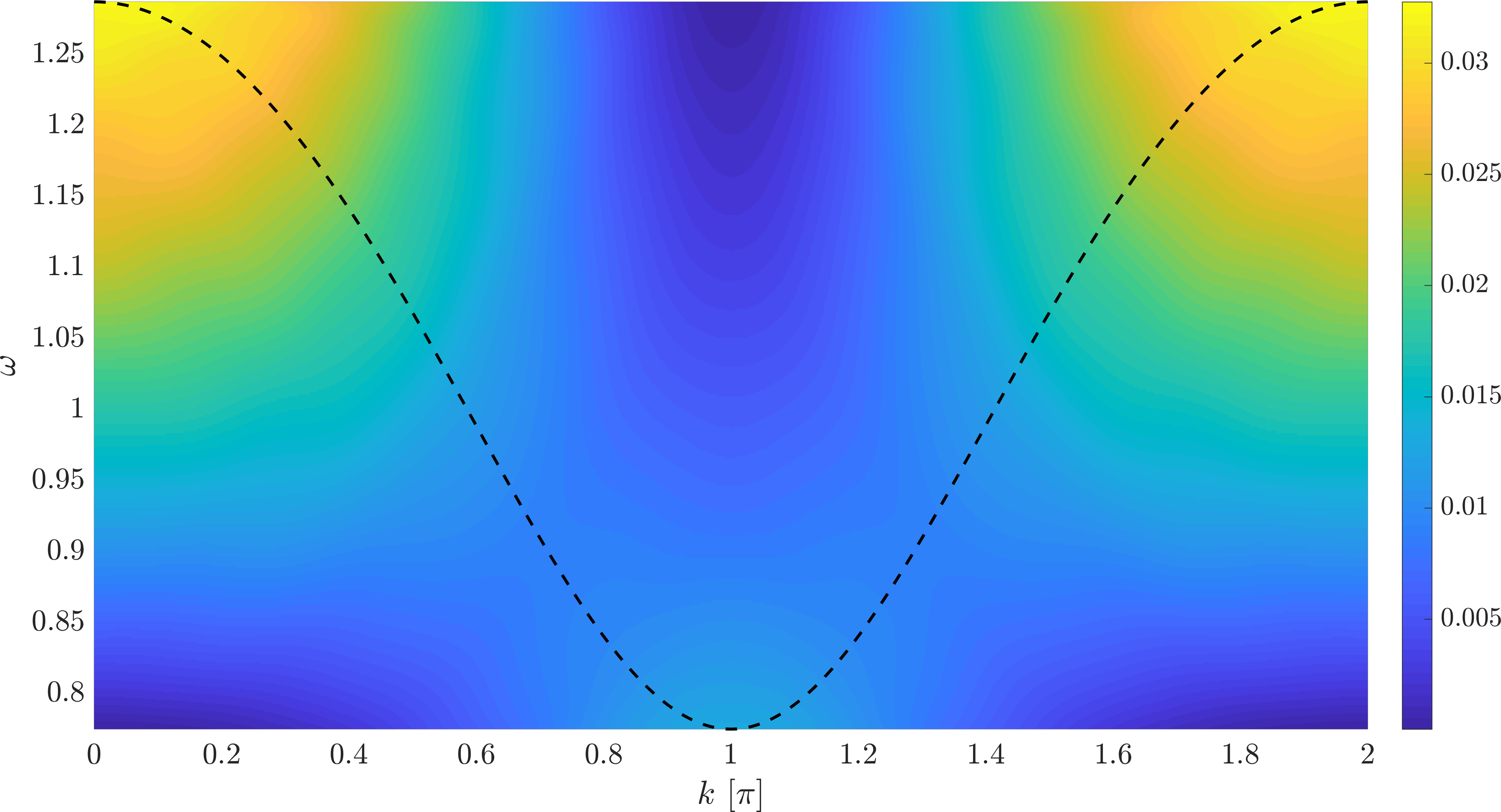}
	\caption{$\Braket{\vert\mathcal{H}_1(k,k'(\omega))\vert}^2+\Braket{\vert\mathcal{H}_1(2\pi-k,k'(\omega))\vert^2}$ is shown as a function of $k$ and $\omega$ for the leg disorder setup of the main body of the paper. Note that $\omega(k')=\Braket{\mathcal{H}_1(k',k')}$. Further at the position of the dashed curve $\Var(\mathcal{H}_1(k,k'(\omega)))+\Braket{\vert\mathcal{H}_1(2\pi-k,k'(\omega))\vert^2}$ is shown. The dashed line depicts the dispersion of $\Braket{\mathcal{H}_1(k,k)}$. }
	\label{DisorderPotential}
\end{figure}

In first-order Born approximation the mean self-energy is 
\begin{equation}
\overline{\Sigma}(k,w)_{\mathrm{Born}}=\int \Braket{\vert V(k,k'\vert^2}G^{0+}_{0}(k',w)\rm{d}k'.
\end{equation}
Here $V(k,k')=\mathcal{H}_1(k,k')$ for $k\neq k'$, $\mathcal{H}_1$ is the one-triplon effective Hamiltonian and $V(k,k)=\mathcal{H}_1(k,k)-\Braket{\mathcal{H}_1(k,k)}$. The retarded unperturbed Green's function is that of the mean hopping amplitudes, i.e.
\begin{equation}
G^{0+}_{0}(k',w)=\frac{1}{w-\Braket{\mathcal{H}_1(k',k')}+\rm{i}0^+}.
\end{equation}
Note that in the thermodynamic limit $\mathcal{H}_1(k',k')=\Braket{\mathcal{H}_1(k',k')}$. For $w=\Braket{\mathcal{H}_1(k',k')}\stackrel{!}{=}\Braket{\mathcal{H}_1(k,k)}$ the Dirac identity yields us two contributions to the imaginary part of the self-energy (assuming that $k=0,\pi$ are the only maxima/minima)
\begin{equation}
\Im\left(\overline{\Sigma}(k,\Braket{\mathcal{H}_1(k,k)})_{\mathrm{Born}}\right)=\left\vert\frac{1}{\frac{d}{dk'}\Braket{\mathcal{H}_1(k',k')}\vert_{k'=k}}\right\vert\left(\Var(\mathcal{H}_1(k,k))+\Braket{\vert\mathcal{H}_1(k,2\pi-k)\vert^2}\right).
\end{equation}
Thus the imaginary part of the self-energy at $(k,\Braket{\mathcal{H}_1(k,k)})$ and the approximate broadening of the momentum state $\Ket{k}$ scales proportionally to \begin{equation}
\Var(\mathcal{H}_1(k,k))+\Braket{\vert\mathcal{H}_1(2\pi-k,k)\vert^2}.
\label{ScalingFactor}
\end{equation} Although $\lvert\frac{1}{\frac{d}{dk'}\Braket{\mathcal{H}_1(k',k')}\vert_{k'=k}}\rvert=0$ the broadening at $k=0,\pi$ is assumed to scale similarly.\\
For increasing $\Braket{J^{\parallel}_\nu}$ there are big differences in the momentum lifetime at $k=0$ and $k=\pi$. This can be seen in the main body of the paper for the leg disorder case in the one-triplon sector (Fig. 2, middle), where momentum states had a significantly longer lifetime at $k=\pi$ than at $k=0$. In Fig.~\ref{DisorderPotential} the scaling factor \eqref{ScalingFactor} for the momentum state lifetimes described above is shown for the same leg disorder setup. The important values are those that coincide with the mean dispersion $\Braket{\mathcal{H}_1(k',k')}$ depicted as dashed line in the figure. Clearly, one recognizes that at $k=\pi$ momentum states will have a significantly longer lifetime than at $k=0$. In the following the matrix elements of $\mathcal{H}_{1,\nu,\nu'}$ are denoted as $a_{\nu,\nu'}$. As cause for the different behaviour at the mean band edges correlations between the local hopping term $a_{\nu,0}$ and other hopping terms were identified. To estimate the importance of the different correlations between hopping elements on the lifetime of momentum states, the following relation proved useful.\\
For correlated but stationary disorder, i.e. $\Braket{J^{\parallel}_{\nu} J^{\parallel}_{\nu+\delta}}\neq 0$, $\Braket{J^{\perp}_{\nu} J^{\perp}_{\nu+\delta}}\neq 0$ and $\Braket{J^{\parallel}_{\nu} J^{\perp}_{\nu+\delta}}\neq 0$ but independent of $\nu$, the mean absolute square of the Hamiltonian in the momentum basis can be given using the cross-correlation theorem. Starting point is $\mathcal{H}_1$ transformed into the momentum basis.
\begin{equation}
\begin{aligned}
\mathcal{H}^{}_{1,k',k}=&\frac{1}{\mathcal{N}_{r}}\Bra{\sum_{\nu'}e^{-{\rm i}\nu'k'}\ket{\nu'}}\mathcal{H}^{}_{1}\Ket{\sum_{\nu}e^{-{\rm i}\nu k}\ket{\nu}}\\
=&\frac{1}{\mathcal{N}_{r}}\delta_{k,k'}\sum_\nu\left(a_{\nu,0}+2\sum_{t=1}^{}a_{\nu,t}\cos(tk)\right)\\
&+\frac{1}{\mathcal{N}_{r}}(1-\delta_{k,k'})\sum_{\nu}e^{{\rm i}\nu(-k+k')}\left(a_{\nu,0}+\sum_{t=1}^{}a_{\nu,t}(e^{{\rm i}tk'}+e^{-{\rm i}tk})\right).
\end{aligned}
\label{CrossCor}
\end{equation} 
All that has to be known to use the cross-correlation theorem is the covariance structure of the hopping amplitude random processes. One obtains
\begin{equation}
\begin{aligned}
&\mathcal{N}_{r}\Braket{\vert \mathcal{H}_{1,k,k'}\vert^2}=2\sum_{d }\sum_{\substack{t,t'=1\\t>t'}}^{}\Cov\left(a_{\nu,t}a_{\nu+d,t'}\right)\bigg(\cos\left((t'-t)k+d(k-k')\right)\\
&+\cos\left((t-t')k'+d(k-k')\right)+\cos\left(t'k+tk'+d(k-k')\right)+\cos\left(-t'k'-tk+d(k-k')\right)\bigg)\\
&+2\sum_{d}^{}\sum_{t=1}^{}\Cov\left(a_{\nu,0}a_{\nu+d,t}\right)\left(\cos\left(tk+d(k-k')\right)+\cos\left(-tk'+d(k-k')\right)\right)\\
&+\sum_{d\geq0}^{}\sum_{t=0}^{}\left(2(1-\delta_{d,0})\cos\left(d(k-k')\right)+\delta_{d,0}\right)\\
&\cdot\left(\delta_{t,0}\Cov\left(a_{\nu,0}a_{\nu+d,0}\right)+(1-\delta_{t,0})\Cov\left(a_{\nu,t}a_{\nu+d,t}\right)2(1+\cos\left(t(k+k')\right)\right).
\end{aligned}
\end{equation}
\\ With this formula, which is also valid in the thermodynamic limit, it is easy to estimate the importance of correlations between hopping elements for the lifetime of momentum states.

\section{Leading-order effects of rung disorder}
\begin{figure}[htbp]
\centering
\includegraphics[width=0.9\columnwidth]{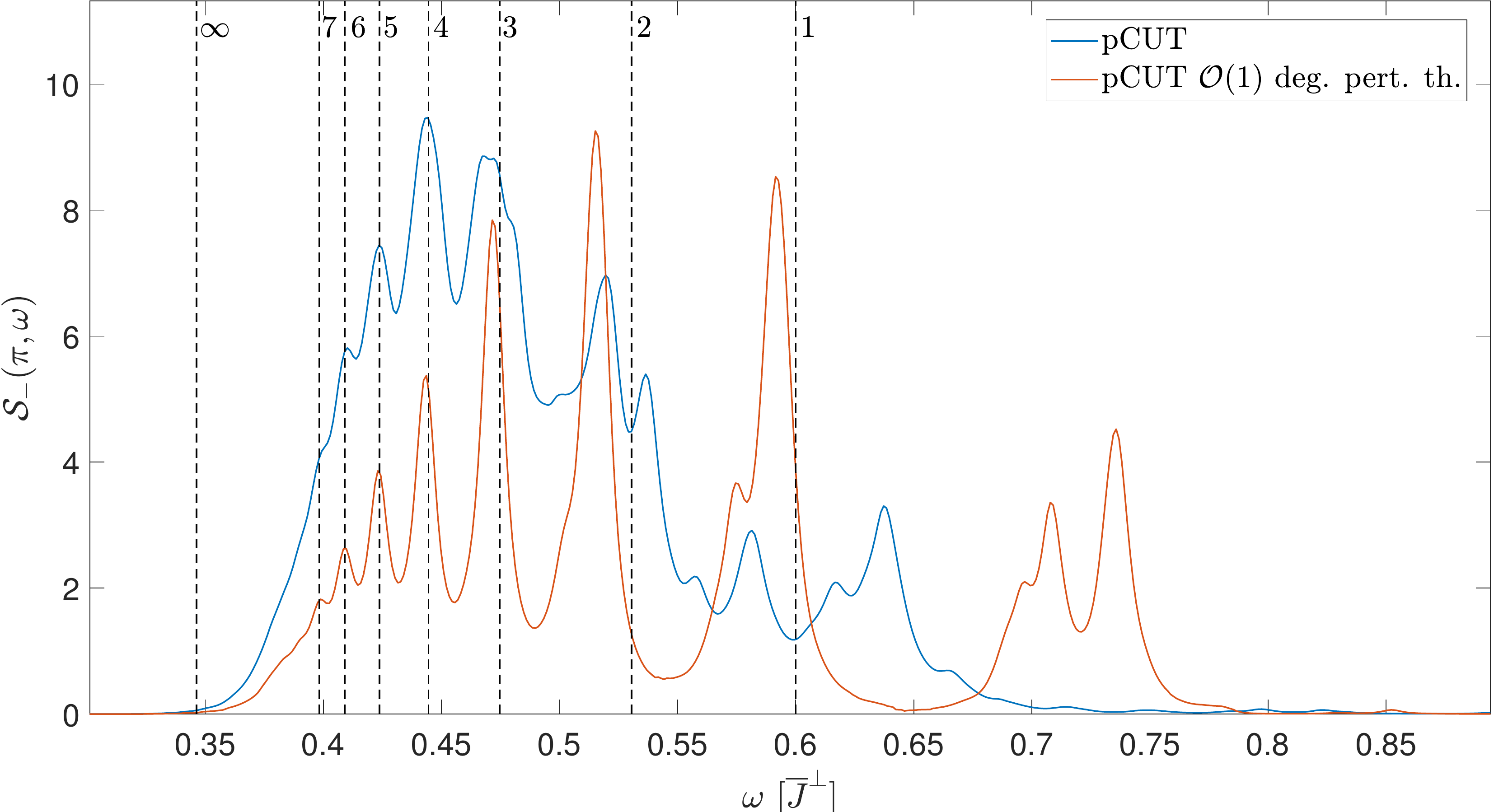}
\caption{In blue the dynamic structure factor at $k=\pi$ is depicted for the rung disorder setup of the main body of the paper with parameters $\mathcal{N}_r=100$, $\mathcal{N}_{\rm{dc}}=1000$ and $\Gamma=0.01$. The red curve shows results by applying first-order degenerate perturbation theory on the one-triplon Hamiltonian used to calculate the blue curve. The vertical black lines indicate the lowest energy $\epsilon_1^{(L)}$ of a ladder segment of $L$ rungs with value $J^\perp=0.6$ and open boundary conditions.}
\label{RungEffects}
\end{figure}

For strong rung disorder one can do degenerate perturbation theory around the unperturbed Hamiltonian of only local triplon interactions. This yields in first order two Hamiltonians without interactions between rungs with different rung couplings. The low-energy physics is described by the Hamiltonian with small rung values and the high-energy physics by the one with the large values. An important issue of this perturbative picture are the lengths of chains of consectutive low or high rung values. The mean amount of such chains of length $L$ is $\mathcal{N}_r(1-p)^2p^L$ if $p$ is the probability for the specific rung value and $\mathcal{N}_r=\infty$. For a model with only nearest-neighbour hopping all the physics in first-order degenerate perturbation theory is happening in those finite ladder segments. When hopping to further neighbours is taken into account or when higher orders in the degenerate perturbation theory are not neglected those finite chains interact with each other. Fig. \ref{RungEffects} shows the dynamic structure factor at $k=\pi$ for the rung disorder setup of the main body of the paper and in the one-triplon sector (blue curve). In red the result obtained with first-order degenerate perturbation theory in the two different rung values applied on the one-triplon Hamiltonian used to calculate the blue curve is shown. The vertical black dashed lines show the smallest energies of consecutive $J^\perp=0.6$ chains of length $L$ with open boundary conditions. For $L=7,6,5,4,(3)$ these black lines match well with both the maxima of the blue and red curves. This indicates that the discrete nature of the length of these ladder segments is the main reason for the appearance of multiple maxima in Fig. \ref{RungEffects} and the approximate quantization of the DSF that was seen in the main body of the paper.\\ The intensities of the red curves are biggest at $L=1,2$. This is what one would find assuming that the weight at $k=\pi$ is proportional to $L$ and neglecting hopping further than to next neighbours. Then the intensities are $\propto L\,\mathcal{N}_r(1-p)^2p^l$ when only the smallest energy $\epsilon_1^{(L)}$ contributes to the DSF at $k=\pi$. There are several reasons why the intensities of the exact case differ. There is a shift towards lower energies caused by the interactions between several finite ladder segments. The blue curve rather consists of peaks on a plateau of constant intensity. It is caused by the splitting of energies induced by these additional interactions. Because more than $L$ rungs can carry the weight of eigenfunctions due to additional interactions neglected in first-order degenerate perturbation theory the weight at $k=\pi$ can scale stronger than with $L$. This is true for all $L$. However, because for a segment of length $L=1$ the eigenstate is a localized rung and thus non-dispersive, there is no preference in the $L=1$-sector for $k=\pi$. In contrast, this is different for larger $L$ and is likely another reason why the blue curve shows stronger intensities there. 

\section{Spectral weights}
\begin{figure}[htbp]
	\centering
	\includegraphics[width=0.9\columnwidth]{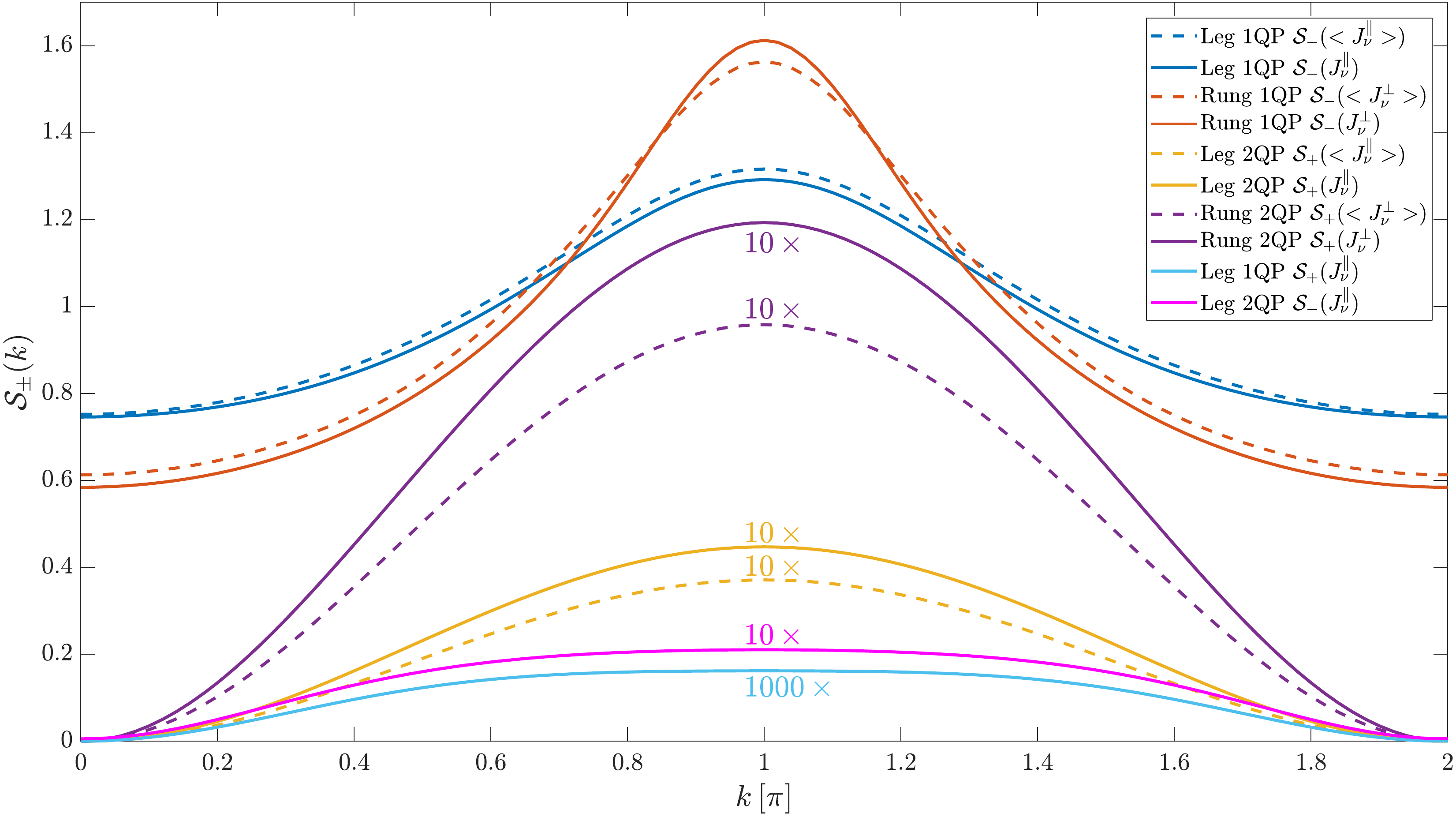}
	\caption{The spectral weights $\mathcal{S}_{\pm}(k)$ are shown as solid lines for all sorts of disorder considered in the paper (denoted as $\mathcal{S}_{\pm}(J_\nu^\perp,J_\nu^\parallel)$). The spectral weights with constant mean rung and leg values are plotted as dashed lines for the same disorder setups (denoted as $\mathcal{S}_{\pm}(\braket{J_\nu^\perp,J_\nu^\parallel)}$).}
	\label{StaticSF}
\end{figure}

Fig.~\ref{StaticSF} shows the spectral weights $\mathcal{S}_{\pm}(k):=\int \mathcal{S}_{\pm}(k,\omega)\rm{d}\omega$ obtained by pCUTs for all disorder setups discussed in the main body of the paper. Interestingly, the spectral weights of the corresponding models with constant mean rung and leg couplings show only minor differences to the disodered one in the one-triplon sector. In the two-triplon sector the shape is similar but the scale is smaller. \\
The observables $\mathcal{S}_{+}(k)$ ($\mathcal{S}_{-}(k)$) in the one- respectively two-triplon sector that only arise due to the broken reflection symmetry around the centerline of the ladder have a very wide maximum around $k=\pi$ showing only small $k$-dependence there. For $k=0$ the weight of these observables is zero. The symmetric observable is zero at $k=0$ because it commutes with the Hamiltonian. The antisymmetric observable in the two-triplon sector also goes to zero at $k=0$. For that, note that total spin-one two-triplon states have odd parity with respect to inversion of the lattice around their center of mass. At the same time $\mathcal{O}_-(0)$ and the singlet reference state have even parity for that symmetry and since these parities are preserved by the pCUT $\mathcal{O}_-(0)$ shows no weigth in that sector.\\
One can conclude that the huge differences in the dynamic structure factor between the disordered and the non-disordered case are dominantly caused by changes in the dynamics and not by the spectral weights.

\section{Density of states}
The density of states (DOS) was calculated numerically for all disorder configurations $\{J\}$ using a bin width of $0.0014$ and averaging over $\mathcal{N}_{\rm{dc}}=1000$ samples of size $\mathcal{N}_{r}=100$. It is normalized such that its integral over energy is $1$. The result is shown in the upper panel of Fig.~\ref{IPRDOS}. The DOS has large influence on the dynamic structure factor for rung disorder in the one-triplon sector. For that disorder many regions with vanishing density of states occur. Also in the non-vanishing energy regions the DOS shows strong oscillatory behaviour for one-triplon rung disorder. For leg disorder the DOS is mainly smooth. At higher energies we see oscillatory behaviour again that corresponds to eigenfunctions centered around $k=0$ where local hopping amplitudes play a larger role.\\
In the two-triplon sector rung disorder does not lead anymore to regions of vanishing DOS. Still, we find an oscillatory behaviour especially at larger energies. In contrast, for leg disorder differences to the non-disordered case become small and we see an almost-smooth DOS.

\section{Inverse participation ratio}

\begin{figure}[htbp]
\begin{center}
\includegraphics[width=1\textwidth]{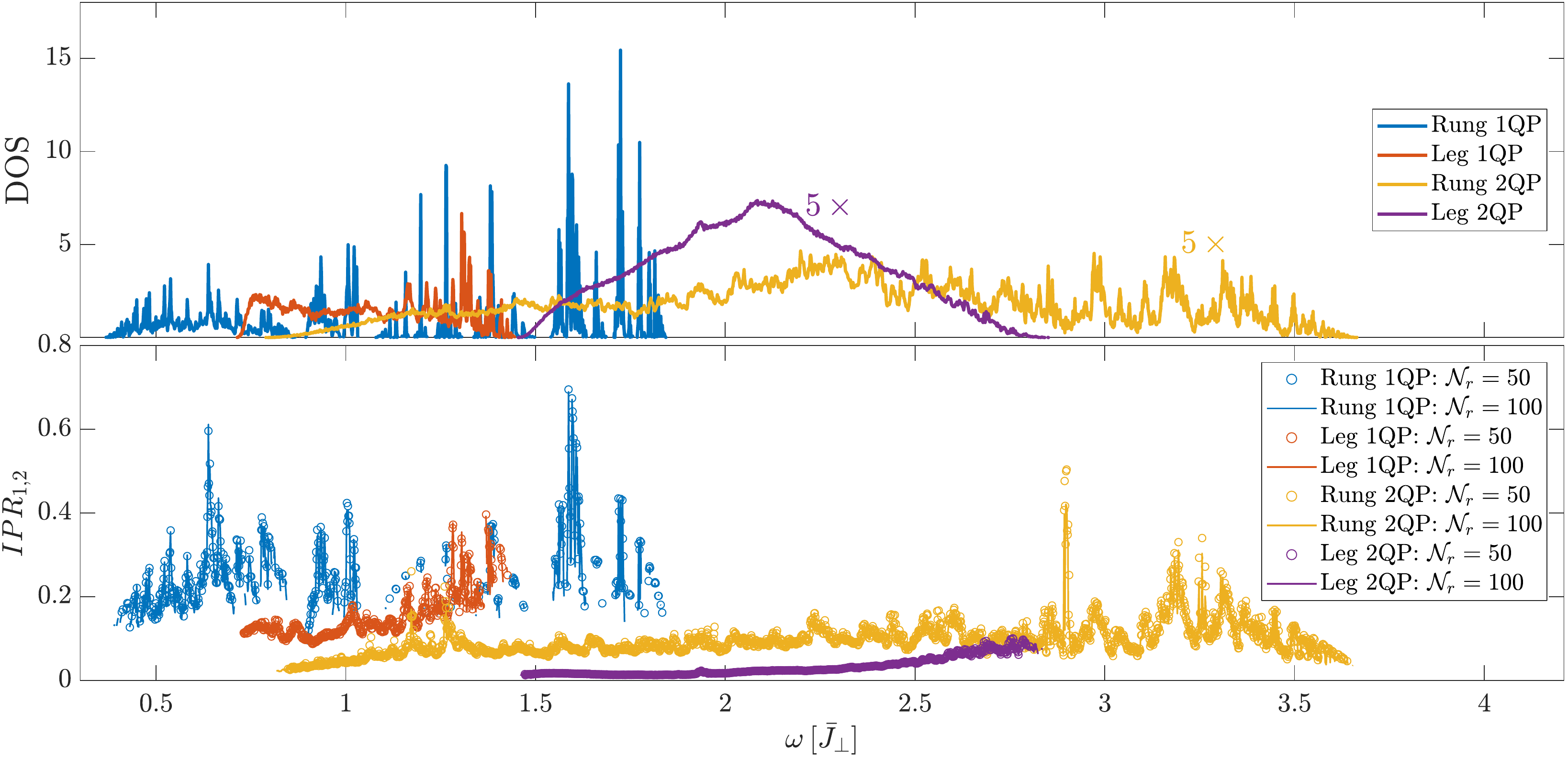}
\caption{Upper panel: the inverse participation ratios $IPR_{1,2}$ are plotted for $\mathcal{N}_{r}=100$ and $\mathcal{N}_{\rm{dc}}=1000$ (solid lines) and  $\mathcal{N}_{r}=50$ and $\mathcal{N}_{\rm{dc}}=2000$ (circles) for all disorder setups used in the main body of the paper. Lower panel: the density of states (DOS), which is normalized with respect to energy, is shown for the same disorder setups and $\mathcal{N}_{r}=100$ and $\mathcal{N}_{\rm{dc}}=1000$. The bin width used to calculate both quantities was $0.0014$.}
\label{IPRDOS}
\end{center}
\end{figure}

The inverse participation ratio ($IPR_1$)
\begin{equation}
IPR_1 = \sum_{\nu} \left\vert \Braket{n\vert\nu}\right\vert^4
\end{equation}
with $\Ket{\nu}$ denoting position states is a simple and intuitive measure for the localization length of a normalized eigenfunction $\ket{n}$. Suppose $\ket{n}$ is a plane wave eigenstate of a periodic one-dimensional chain with one atom in the unit cell. Then the $IPR_1$ will be $1/\mathcal{N}_{r}$, $\mathcal{N}_{r}$ being the length of the chain. For all other extended states the IPR will be bigger but always remain $\propto 1/\mathcal{N}_{r}$ \cite{IPR}. For a perfectly localized eigenstate with $\Ket{n}=\Ket{\nu}$ the value of the $IPR_1$ is one. An exponentially localized state 
\begin{equation}
\left\vert \Braket{n\vert\nu}\right\vert ^2 \propto \exp\left(-\frac{\vert \nu - \nu_0\vert}{\xi}\right)
\end{equation}
has an $IPR_1$ of 

\begin{equation}
\frac{\sum_{\nu}\exp\left(-2\frac{\vert \nu - \nu_0\vert}{\xi}\right)}{\left(\sum_{\nu}\exp\left(-\frac{\vert \nu - \nu_0\vert}{\xi}\right)\right)^2} = \frac{2\;\frac{1}{1-\exp\left(-\frac{2}{\xi}\right)}-1}{\left(2\;\frac{1}{1-\exp\left(-\frac{1}{\xi}\right)}-1\right)^2}\approx\frac{1}{4\xi}.
\end{equation}
The approximation used in the last step is already quite good for localization lengths $\xi\geq1$. With that the localization length can be estimated as $\xi \approx \nicefrac{1}{(4IPR_1)}$. The $IPR_1$ ranges between $1$ for a perfectly localized state and $0$ for extended states as the system size $\mathcal{N}_{r}\rightarrow \infty$.\\
For two-particle position states $\Ket{\nu,\nu+\delta}$ and their corresponding eigenfunctions $\Ket{n}$ in position space we use an analogous quantity to $IPR_1$:
\begin{equation}
IPR_2 = \sum_{\nu,\delta} \left\vert \Braket{n\vert \nu,\nu+\delta}\right\vert^4
\end{equation}
Clearly $IPR_2$ goes to zero with system size for extended states as well and only remains finite if the eigenfunctions are localized on a finite subset of two-particle position states. \\
The inverse participation ratio and its two-particle generalization are plotted for $\mathcal{N}_{r}=100$ and $\mathcal{N}_{\rm{dc}}=1000$ (solid lines) and  $\mathcal{N}_{r}=50$ and $\mathcal{N}_{\rm{dc}}=2000$ (circles) in the lower panel of Fig.~\ref{IPRDOS}. In both cases a bin width of $0.0014$ was used.
There are only small deviations when the number of rungs gets changed from $\mathcal{N}_{r} = 100$ to $\mathcal{N}_{r} = 50$. The conclusion is that the energy states fit
with almost all their weight already on $50$ rungs and that there should be hardly any finite-size effects
anymore for systems of $100$ rungs or more.\\
For leg disorder the $IPR_{1,2}$ increases towards
higher energies. These have biggest weight on momentum $k=0$. Higher order effects lead
to stronger broadening at $k=0$ and this explains why the $IPR_{1,2}$ grows towards higher
energies.\\
The $IPR_2$ in the two-triplon sector shows essentially the same features as the $IPR_1$ in the one-triplon sector. We stress that it remains finite so that also two-triplon states have almost all their weight on a finite part of the two-triplon space in the position basis. One main difference to the $IPR_1$ is that
it varies less strongly for the bimodal rung distribution.\\
The average $IPR_{1,2}$ is bigger in the rung disorder case. 

\section{Finite-size effects, convergence of perturbative expansion and statistical error}
As the dynamic structure factor could only be calculated for finite systems possible
finite-size effects have to be examined. Further the convergence of the perturbative expansion has to be checked. Due to the finite number of samples $\mathcal{N}_{\rm{dc}}$ a statistical error arises which has to be quantified.
\begin{figure}[b]
	\centering
	\includegraphics[width=1\columnwidth]{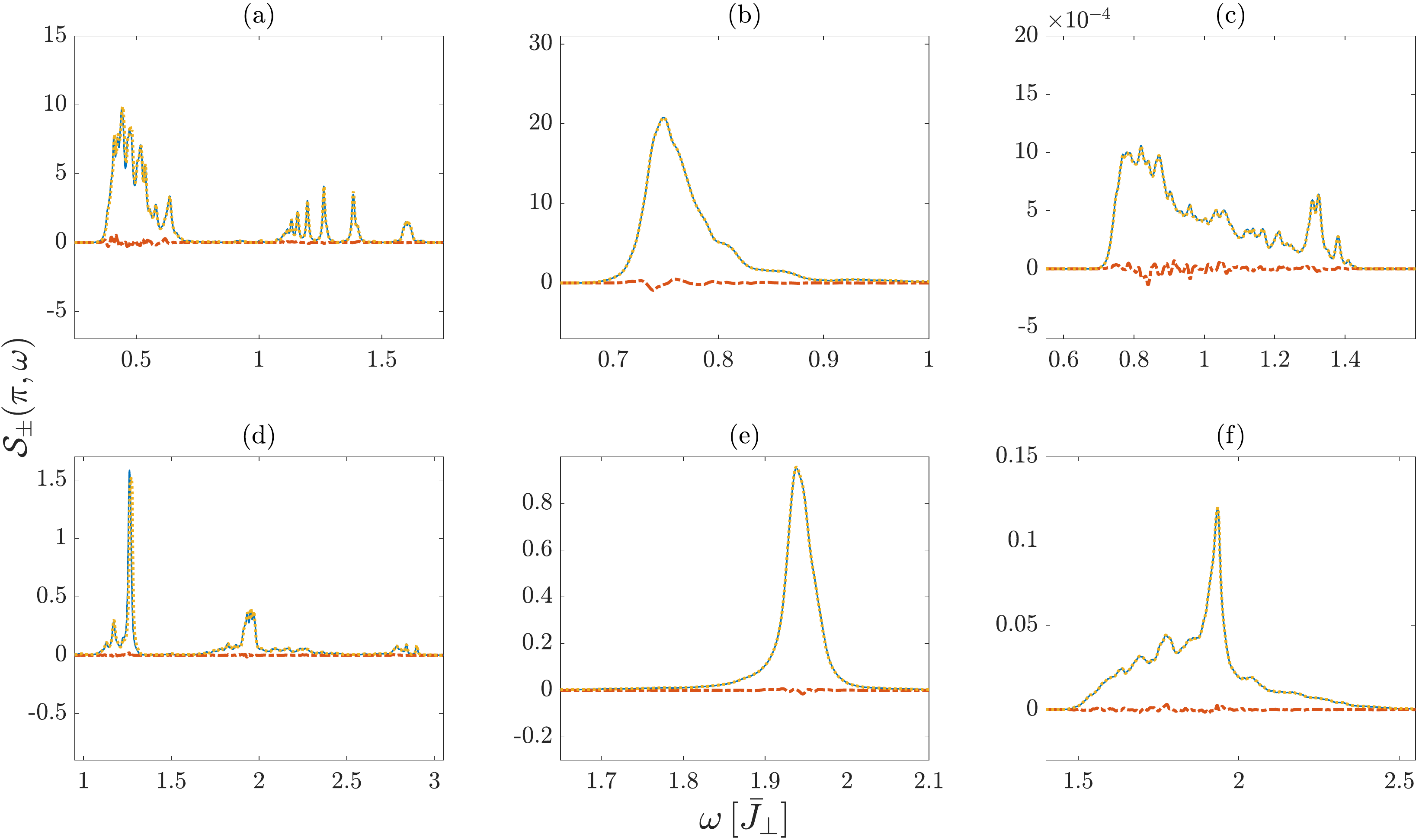}
	\caption{In blue the dynamic structure factor at $k=\pi$ is shown for $\mathcal{N}_{\rm{dc}}=100$ samples of $\mathcal{N}_{r}=100$. The red dashed lines display the difference of these values and the ones obtained with $\mathcal{N}_{\rm{dc}}=200$ samples of $\mathcal{N}_{r}=50$ but with the same disorder samples. Yellow dotted lines show the dynamic structure factor for the same disorder samples and $\mathcal{N}_{r}=100$ but with order $7$ in the perturbative expansion of the matrix elements of the Hamiltonian and order $6$ for the observables only. The same disorder setups as in the main body of the paper were chosen and the same Lorentzian curve for broadening was used ($\Gamma=0.01$). One-triplon sector: (a) rung disorder $\mathcal{S}_{-}$, (b) leg disorder $\mathcal{S}_{-}$, (c) leg disorder $\mathcal{S}_{+}$. Two-triplon sector: (d) rung disorder $\mathcal{S}_{+}$, (e) leg disorder $\mathcal{S}_{+}$, (f) leg disorder $\mathcal{S}_{-}$.}
	\label{NConv}
\end{figure}
\\In Fig.~\ref{NConv} the dynamic structure factor at $k=\pi$ is shown for all disorder setups of the main body of the paper. The blue line shows results for $\mathcal{N}_{\rm{dc}}=100$ and $\mathcal{N}_{r}=100$. The yellow dotted curve was calculated with the same disorder samples but with order $7$ in the perturbative expansion of matrix elements of the Hamiltonian and order $6$ for the observables only. For leg disorder hardly any differences can be made out. Thus the perturbative expansion is well converged there. For rung disorder at the peaks differences can be seen which are due to a small offset in the energy of the peaks between the order $8$ and $7$ calculation. Anyway these offsets are quite small and the convergence of perturbation theory is also expected to be good for rung disorder. \\
Fig.~\ref{NConv} further shows the difference of the calculation with $\mathcal{N}_{\rm{dc}}=200$ and $\mathcal{N}_{r}=50$ and $\mathcal{N}_{\rm{dc}}=100$ and $\mathcal{N}_{r}=100$ by using the same disorder samples again (red curve). For that each sample of the $\mathcal{N}_{r}=100$ calculation was cut into two samples of $\mathcal{N}_{r}=50$. The deviations found are quite small and biggest in the one-triplon leg disorder setups. 
\\
The calculation can only be done for a finite number of
samples and a finite width of energy bins or by folding with a Lorentzian of finite width $\Gamma$.
Assuming that the value in each bin is a random variable and that the realizations of disorder
are independent from sample to sample the behaviour of its standard deviation with
the number of samples $\mathcal{N}_{\rm{dc}}$, the number of rungs in that samples $\mathcal{N}_{r}$ and broadening $\Gamma$ of the Lorentzian is $\propto 1/\sqrt{\mathcal{N}_{\rm{dc}}\,\mathcal{N}_{r}\,\Gamma}$.\\
Fig.~\ref{NRConv} shows the relative standard deviation $\sigma\,[\mathcal{S}_\pm(\pi,\omega)]/\Braket{\,\mathcal{S}_\pm(\pi,\omega)\,}$ for all disorder setups used in the main body of the paper and for parameters $\mathcal{N}_{r}=100$ and $\mathcal{N}_{\rm{dc}}=1000$ (blue circles). Data points where the mean values were smaller than $1/10$ of the maximum value were excluded. One can see that the statistical error is smaller than $5\%$ for those values. 

\begin{figure}[htbp]
	\centering
	\includegraphics[width=1\columnwidth]{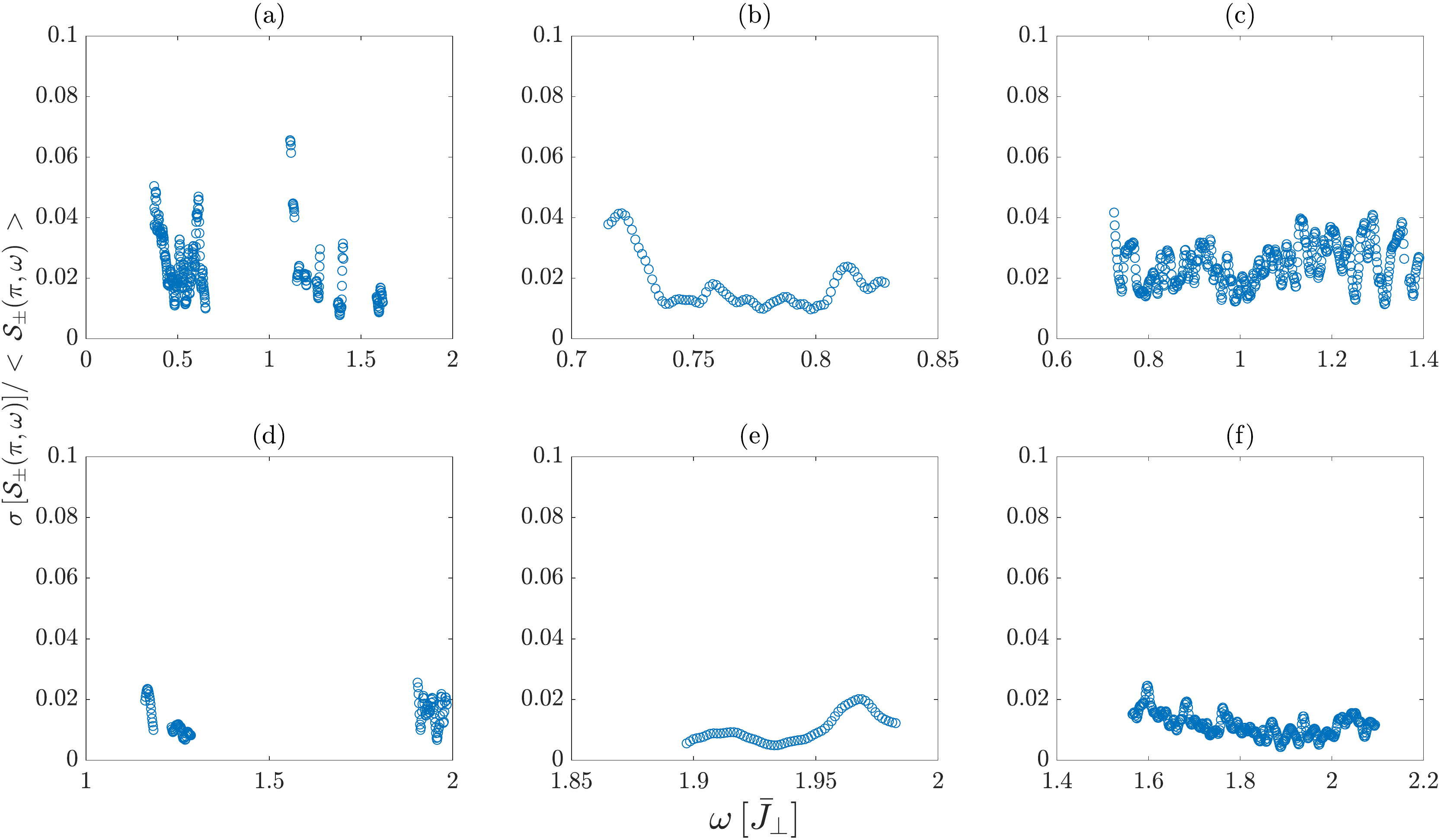}
	\caption{In blue the relative standard deviation $\sigma\,[\mathcal{S}_\pm(\pi,\omega)]/\Braket{\,\mathcal{S}_\pm(\pi,\omega)\,}$ of systems with $\mathcal{N}_{r}=100$ and $\mathcal{N}_{\rm{dc}}=1000$ is shown. The same disorder setups as in the main body of the paper were chosen and the same Lorentzian curve for broadening was used ($\Gamma=0.01$). Data points where the mean values were smaller than $1/10$ of the maximum value were excluded. One-triplon sector: (a) rung disorder $\mathcal{S}_{-}$, (b) leg disorder $\mathcal{S}_{-}$, (c) leg disorder $\mathcal{S}_{+}$. Two-triplon sector: (d) rung disorder $\mathcal{S}_{+}$, (e) leg disorder $\mathcal{S}_{+}$, (f) leg disorder $\mathcal{S}_{-}$.}
	\label{NRConv}
\end{figure}

\bibliographystyle{apsrev4-1}


\end{document}